\def\year{2021}\relax
\documentclass[letterpaper]{article} 
\usepackage{aaai21}  
\usepackage{times}  
\usepackage{helvet} 
\usepackage{courier}  
\usepackage[hyphens]{url}  
\usepackage{graphicx} 
\usepackage{natbib}  
\usepackage{caption} 
\usepackage{comment}

\title{VaccinEU: COVID-19 vaccine conversations on Twitter \\ in French, German and Italian}
\author{Marco Di Giovanni\textsuperscript{\rm 1, 3,*}, Francesco Pierri\textsuperscript{\rm 1,2,*} Christopher Torres-Lugo\textsuperscript{\rm 2} and Marco Brambilla\textsuperscript{\rm 1}
\\}

\affiliations{
    \textsuperscript{\rm 1}Dipartimento di Elettronica, Informazione e Bioingegneria, Politecnico di Milano, Italy\\
    \textsuperscript{\rm 2}Observatory on Social Media, Indiana University, Bloomington, USA\\
    \textsuperscript{\rm 3}Università di Bologna, Italy\\
    \textsuperscript{\rm *}These authors contributed equally to this work.\\
    Corresponding author: francesco.pierri@polimi.it\\
}

\begin{document}

\maketitle

\begin{abstract}
Despite the increasing limitations for unvaccinated people, in many European countries there is still a non-negligible fraction of individuals who refuse to get vaccinated against SARS-CoV-2, undermining governmental efforts to eradicate the virus. We study the role of online social media in influencing individuals' opinion towards getting vaccinated by designing a large-scale collection of Twitter messages in three different languages -- French, German and Italian -- and providing public access to the data collected. Focusing on the European context, our VaccinEU dataset aims to help researchers to better understand the impact of online (mis)information about vaccines and design more accurate communication strategies to maximize vaccination coverage. Data can be fully accessed in a Dataverse repository\footnote{ \url{https://doi.org/10.7910/DVN/NZUMZG}} and a GitHub repository\footnote{ \url{https://github.com/DataSciencePolimi/VaccinEU}}.
\end{abstract}

\section{Introduction}
Less than a year into the COVID-19 pandemic, the first vaccine was approved and made available to the public\footnote{\url{https://www.fda.gov/news-events/press-announcements/fda-approves-first-covid-19-vaccine}}, providing an effective tool to fight the spread of the virus \cite{orenstein2017simply}. Vaccination programs started towards the end of 2020 in most European countries, and as of December 2021 over 700 M doses have been administered according to Our World in Data\footnote{https://ourworldindata.org/covid-vaccinations}. However, despite the large availability of vaccines, vaccine uptake exhibits a large variability across different countries, ranging from 40\% of people vaccinated with at least one dose in Romania to 90\% in Portugal\footnote{https://vaccinetracker.ecdc.europa.eu/public/extensions/COVID-19/vaccine-tracker.html}. This indicates that a considerable number of people are still hesitant to get vaccinated, and that it will be hard to reach herd immunity . 

Research in the past highlighted the role of online social media in promoting and amplifying negative views about vaccines \cite{burki2019vaccine,broniatowski2018weaponized,johnson2020online}. Specifically to the COVID-19 pandemic, concern has recently risen around the 'infodemic' \cite{zarocostas2020fight,yang2020covid,gallotti2020assessing} of misleading information about the virus spreading online, and it has been shown that online misinformation might negatively influence individuals' opinion towards getting vaccinated \cite{pierri2021impact,loomba2021measuring}.

In this paper, we describe a data resource which will allow researchers and academics to study the impact of online conversations about COVID-19 vaccines on Twitter in three different languages: French, German, and Italian. 

Specifically to the Italian context, \citet{righetti_2020} and \citet{cossard_2020} analyzed the debate on Twitter around the 2017 mandatory child vaccination law, observing the spread of problematic information and highlighting the presence of echo chamber effects \cite{cinelli2021echo}.
~\citet{Gargiulo2020Asymmetric} obtained similar results when analyzing French data, finding that defenders and critics of vaccines focus on different topics, and that, while there are more defenders, critics are more active and coordinated. 
To the best of our knowledge, there is no previous work which analyzes vaccine conversations on social media in German langauge.

Our contribution is manifold. We curated a list of vaccine-related keywords as complete as possible with the help of native speakers, using a snowball sampling approach \cite{deverna2021covaxxy}, and collected over 70 million tweets in three different languages, from November 1st 2020 to November 15th 2021, using a combination of streaming and historical search Twitter APIs. To the best of our knowledge there are no such datasets publicly available, with the only exception of VaccinItaly \cite{pierri2021vaccinitaly} in Italian language. We provide public access to this data in agreement with Twitter terms of service by releasing \textit{ids} of tweets which can be used to retrieve full objects via APIs. For each language, we further collected a list of hashtags which strongly state a stance in favor or against vaccination, and we manually annotated a random sample of 1,000 tweets with four labels (Pro-vaccines, Anti-Vaccines, Neutral, Out-of-context). We provide full access to this metadata, which can be used to better understand the polarized debate around vaccinations and train machine learning classifiers to automatically detect anti-vaccination messages \cite{digiovanni2021content}. Finally, we provide some preliminary analyses of the dataset in terms of volumes, hashtags, sources, geolocation and coordinated activity.

The outline of the paper is the following: we first overview existing datasets which relate to our work. Then, we describe in detail the data collection process. Next, we provide some preliminary analyses of the data, leaving more sophisticated analyses for future work. Finally, we discuss limitations and potential uses of this dataset.


\section{Related datasets}
Here we describe some public data resources recently released to study conversations around COVID-19 vaccines on social media.

At the beginning of 2021, \citet{deverna2021covaxxy} released the first Twitter dataset conceived to investigate English language online conversations around COVID-19 vaccines. They used a snowball sampling approach to curate a list as complete as possible of terms related to vaccines, and they provide public access to \textit{ids} of tweets collected since the beginning of January 2020. They also have an associated online dashboard (CoVaxxy\footnote{https://osome.iu.edu/tools/covaxxy}), where they provide an interactive visualization of the relationship between online misinformation spreading on Twitter and the evolution of the US vaccination program. Associations between online misinformation and vaccine hesitancy were reported in \citet{pierri2021impact} leveraging their data.

\citet{pierri2021vaccinitaly} released a public dataset of Italian language tweets related to vaccines and collected since December 2020 to October 2021\footnote{https://github.com/frapierri/VaccinItaly}. They also set-up a collection of public posts about vaccines shared by public Facebook pages and groups and gathered through Crowdtangle. Similar to CoVaxxy, they provide an online dashboard where they show visualizations of the interplay between Twitter conversations and the vaccination program in Italy\footnote{http://genomic.elet.polimi.it/vaccinitaly/}.

\citet{muric2021} focused on antivaccine narratives on Twitter and publicly released two data collections, one streaming keyword–centered with more than 1.8 million tweets, and another historical account–level collection with more than 135 million tweets. Both collections are based on English language keywords. They showed that Twitter users who engaged the most in antivaccination narratives are politically right-wing leaning, and that questionable news sources are very active in promoting negative views about vaccines.

\citet{HAYAWI2021} focused on online misinformation around COVID-19 vaccines. After collecting over 15 million tweets, they manually labeled a sample of 15k tweets with the help of medical experts in order to identify unsubstantiated claims and misleading information about vaccines. They eventually trained and test machine learning classifiers on these tweets, reaching up to 98\% of F1-score in the task of classifying vaccine misinformation.

In addition to the aforementioned resources, several datasets have been released to study the COVID-19 pandemic on Twitter, providing oftentimes useful metadata (geolocation, sentiment, gender, etc) in addition to raw tweet \textit{ids} \cite{banda2021large,chen2020tracking, lopez2021augmented,imran2021tbcov}.



\section{Data collection}

\begin{figure}[!t]
    \centering
    \includegraphics[width=\linewidth]{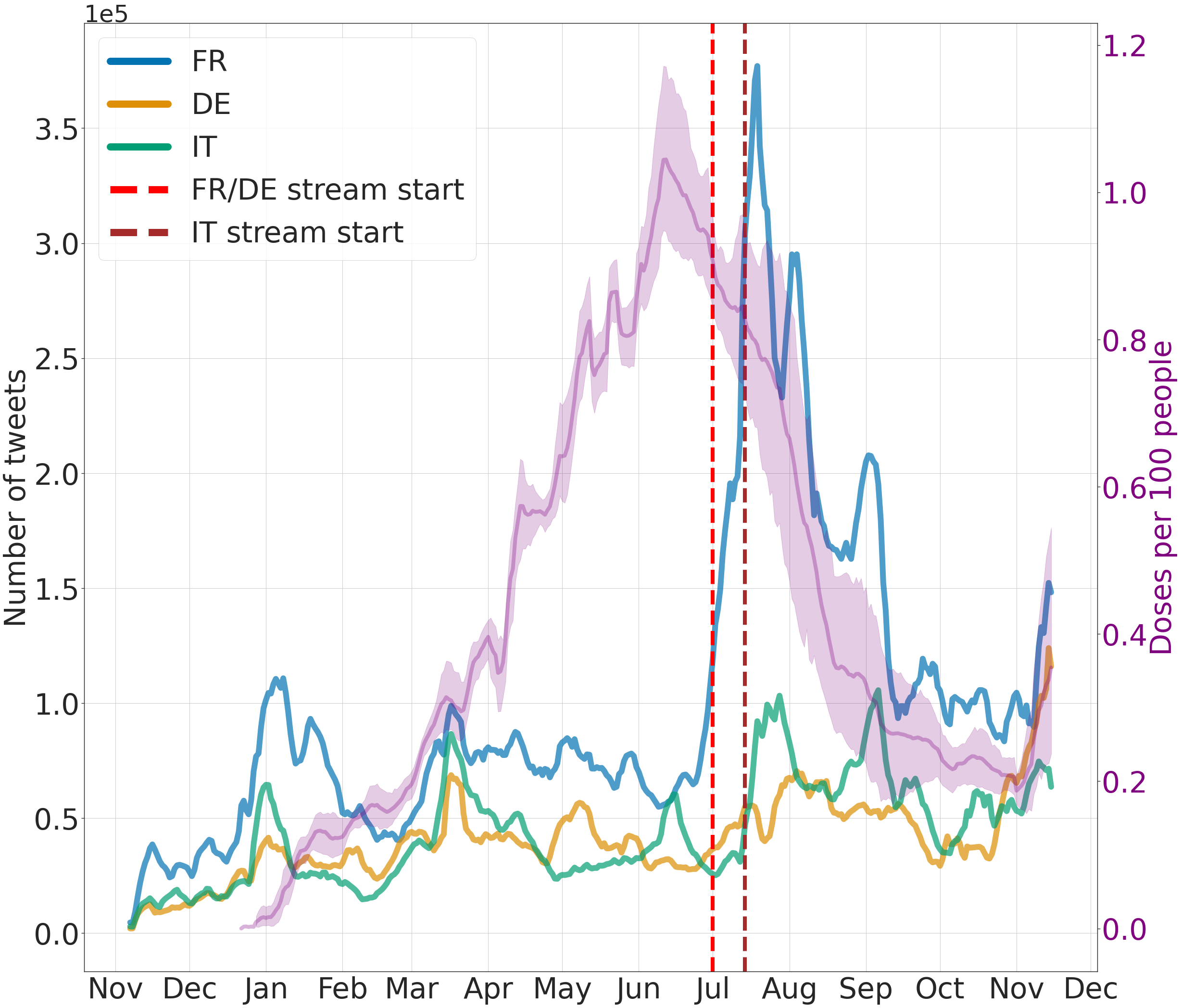}
    \caption{Daily number of vaccine-related tweets collected in different languages (left y-axis), along with the daily number of doses administered per million population (right y-axis) in several European countries (Austria, Belgium, France, Germany and Italy). All time series are smoothed with a 7-day average. Daily vaccinations are obtained from Our World in Data \cite{owid} and correspond to the average over different countries with 95\% C.I. Vertical dashed lines indicate the beginning of the streaming collection for France and Germany (red) and Italy (brown).}
    \label{fig:tweet_vax_ts}
\end{figure}

In this section, we describe our data collection process. 
We detail every design choice made to obtain a dataset as complete and unbiased as possible.

\subsection{Twitter APIs}


We use both the standard streaming Filter API v1.1\footnote{https://developer.twitter.com/en/docs/twitter-api/v1/tweets/filter-realtime/overview} and the new historical Search API v2\footnote{https://developer.twitter.com/en/docs/twitter-api/tweets/search/introduction} to collect tweets related to vaccines in three different languages: French, German, and Italian. 

The Filter API filters tweets that match a defined query in a real-time fashion, up to 1\% of the global stream. Approximately 500 million tweets are shared every day on Twitter\footnote{https://www.internetlivestats.com/twitter-statistics/}, and as shown in Figure \ref{fig:tweet_vax_ts} we collected at most 350k tweets in a day, thus we likely never incur in this limitation. We started the streaming collection of German and French tweets on July 1st, 2021 and Italian tweets on July 14th, 2021. 
We filtered tweets by language specifying the \textit{lang} parameter in the queries. 

We experienced network malfunctioning issues in some cases, and to fill them we used the Historical Search API, which was released at the beginning of 2021, that allows academics and researchers to perform a full-archive search with a set of selected keywords. We also employed it to recover all tweets shared since November 1st, 2020 to June 30th, 2021 (N.B. July 13th for the Italian language). 

\begin{figure}[!t]
    \centering
    \includegraphics[width=\linewidth]{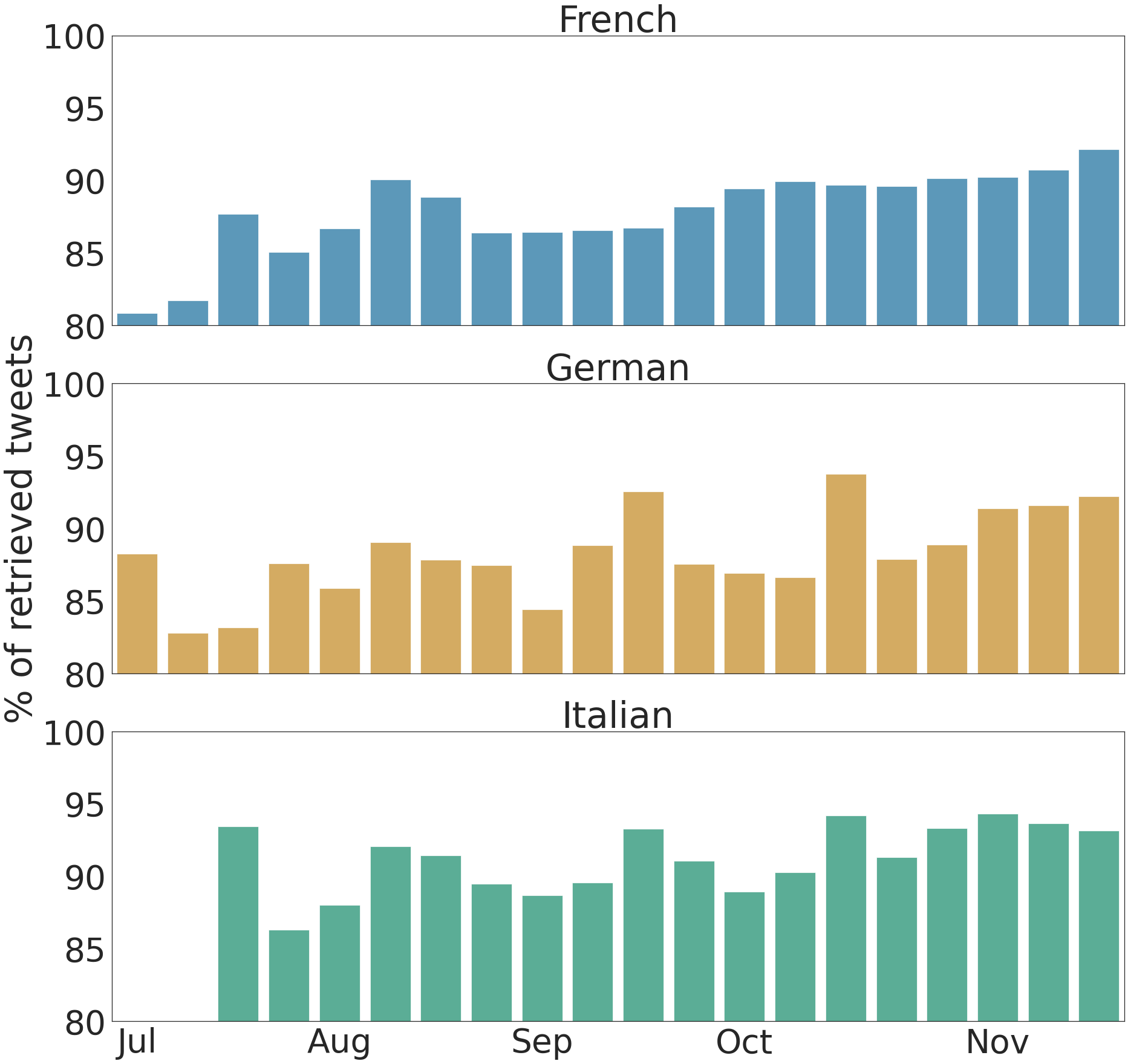}
    \caption{Percentage of tweets successfully retrieved using the \textit{GET statuses/lookup} endpoint. Each point corresponds to a different week, for which we extract a random sample of 10k tweets which we attempt to retrieve. The procedure was done on December 16th, 2021.}
    \label{fig:hydrated_ts}
\end{figure}

We remark that data collected through the historical Search is not complete, due to Twitter's Terms of Service. Twitter does not allow to retrieve deleted tweets nor those shared by protected or suspended accounts\footnote{As a matter of fact, the streaming API also does not provide tweets which are shared by protected accounts.}. Nevertheless, we believe that it is still useful to obtain a collection of vaccine-related tweets as complete as possible. To provide a rough estimate of the amount of tweets that we might lose in the process, we hydrate a random selection of 10k tweets per week collected with the streaming API. We show the percentage of tweets recovered running the \textit{GET statuses/lookup} endpoint on December 16th, 2021 in Figure \ref{fig:hydrated_ts}. We can see that we lost between 5 and 20\% of shared tweets, and that this number likely increases as we search farther in the past.

\subsection{Query Keywords}
Both Filter and Search APIs require one or more keywords to collect relevant tweets. 
An accurate selection of keyword is crucial to obtain a comprehensive dataset. 

We iteratively selected the keywords with the help of three native speakers for each language using a snowball sampling approach \cite{deverna2021covaxxy}. 
We selected as initial set of keywords the translation of very generic vaccine-related words such as "vaccine" and "vaccination" in French, German, and Italian. We made sure to include every grammatically correct variation of words since Twitter APIs perform case-independent \textit{exact} match of keywords and the tokenized texts of tweets (e.g., the tweet "Vaccines are necessary." will be selected if we include in our query the keyword "vaccines", but it will not be collected when including the keyword "vaccine"). 
This might be problematic for languages like German, where words can appear with four different cases (nominative, accusative, dative, and genitive). 

At each round, we used the historical API to filter tweets in the entire period November 2020 - June 2021, and we inspected the most frequent co-occurring words with those in the query. Then, we augmented our list of keywords with those clearly related to vaccines, including specific hashtags, as indicated by native speakers. For instance, we include "\#Igetvaccinated" because tweets containing this hashtag will not be collected by simply using "vaccinated" as keyword.

The final list of keywords for each language is available in our Dataverse\footnote{\url{https://doi.org/10.7910/DVN/NZUMZG}}.

\subsection{Gold Hashtags}
The goal of our project is to understand the influence of positive and negative opinions about vaccines shared on Twitter. 
To this aim, we collected sets of hashtags that indicate the stance (Pro or Anti vaccines) of tweets with high likelihood. We define them as Gold Hashtags (GH), and similarly to our query keywords, we used a snowball sampling approach to obtain a set of hashtags for each language with the help of annotators.
We assume that tweets sharing one or more GH from the same stance express that specific view about vaccines, but this might not always hold true. 

We begun with the selection of one GH for each stance, respectively the translation in different languages of "Iwillgetvaccinated" for Pro and "Iwillnotgetvaccinated" for Anti\footnote{We checked that these hashtags were actually shared by Twitter users in each language.}. 
We iteratively added new GHs inspecting those that co-occurred the most with the initial set of hashtags, based on whether they clearly expressed a stance on vaccines. 
We discarded hashtags when they generically referred to the topic of vaccines, but whose stance was unidentifiable (such as \textit{\#vaccine}). 
We also discarded hashtags that, although their stance seemed clear to the annotators, highly co-occurred with GH of both stances. 
We iterated this procedure three times. The final list of hashtags is available in our repository.


Table~\ref{tab:GHStatistics} shows statistics of GHs. 
Manually inspecting a small set of tweets which included both a Pro and Anti GH, we noticed that most often they do not state a clear stance and usually include questions and pools. 

\begin{table}[t] 
\centering 
\resizebox{.75\columnwidth}{!}{ 
\begin{tabular}{c|ccc}
\textbf{Gold Hashtags} & \textbf{French} & \textbf{German} & \textbf{Italian} \\
\hline
Pro   & 161,871 & 41,933 & 53,374 \\
Anti  & 129,926 & 115,512 & 83,097 \\
Both      & 585 & 1,224 & 451 \\
\hline
\end{tabular}}
\caption{Statistics of tweets sharing Gold Hashtags. }
\label{tab:GHStatistics}
\end{table}

\subsection{Gold Labels}
In addition to hashtags which express a specific stance towards vaccines, we asked our native speakers to manually annotate a sample of random tweets. We randomly picked 1,000 unique tweets for each language, thus discarding retweets, and we asked two annotators to attach one of four "Gold Labels": Pro-vaccines, Anti-vaccines, Neutral, Out-of-Context. We gave them the following guidelines:
Pro- and Anti-vaccines tweets should clearly express a stance about vaccines; Neutral tweets should not express any stance, or their stance is unclear; finally Out-of-Context tweets are tweets not related to COVID-19 vaccines (e.g., animal vaccines). 
A third annotator solved the conflicts by picking one of the two labels for the tweets when they did not agree. 
We report statistics of the labels in Table~\ref{tab:LabelsStatistics}.

\begin{table}[t] 
\centering 
\resizebox{.75\columnwidth}{!}{ 
\begin{tabular}{c|ccc}
\textbf{Gold Labels} & \textbf{French} & \textbf{German} & \textbf{Italian} \\
\hline
Pro-Vaccines   & 419 & 547 & 314 \\
Anti-Vaccines  & 135 & 108 & 151 \\
Neutral        & 279 & 169 & 458 \\
Out-of-Context & 167 & 176 & 77 \\
\hline
\end{tabular}}
\caption{Statistics of manually annotated tweets. }
\label{tab:LabelsStatistics}
\end{table}


\subsection{Data availability}
In agreement with Twitter terms of service, we provide public access to the entire list of tweet \textit{ids} in our Dataverse dataset\footnote{\url{https://doi.org/10.7910/DVN/NZUMZG}} and Github repository\footnote{\url{https://github.com/DataSciencePolimi/VaccinEU}}. These can be "hydrated", i.e., fully retrieved using the \textit{GET statuses/lookup} endpoint of Twitter API, unless they were deleted or their author suspended in the meantime.

In addition to the raw list of \textit{ids}, organized in daily files, we provide the list of \textit{ids} of tweets which contain Pro and Anti vaccine Gold Hashtags (as defined in previous subsection). We also provide the text of tweets labelled using the four Gold Labels defined in the previous subsection.

\section{Data characterization}
In this section we provide descriptive statistics of the data collected in terms of volumes, hashtags, news sources and geolocation. These should be seen as potential uses of the dataset, whereas we leave more sophisticated analyses for future work. In Table \ref{tab:breakdown} we provide basic statistics of the data in terms of tweets, users and URLs for each language.

\begin{table}[!t]
\centering
\begin{tabular}{l|lll}
 & \textbf{French} & \textbf{German} & \textbf{Italian} \\ \hline
Tweets & 38,198,048 & 15,573,108 & 16,581,210 \\
Users & 1,586,071 & 615,317 & 656,578 \\
URLs & 4,749,359 & 2,808,657 & 2,686,055 \\ \hline
\end{tabular}
\caption{Breakdown of the datasets in terms of unique tweets, users and URLs shared, for each language.}
\label{tab:breakdown}
\end{table}

\subsection{Volumes}
In Figure \ref{fig:tweet_vax_ts} we show the daily number of tweets collected for each language, highlighting with two vertical lines when the streaming collection starts for French and German (July 1st 2021), and for Italian (July 14th 2021). As a reference for the COVID-19 vaccination programs, we show the daily number of vaccine doses administered (per 100 people) \cite{owid} averaged over different European countries where these languages are spoken, namely Austria, Belgium, France, Germany, and Italy.

We can see that overall the daily volume of French tweets is much higher compared to the other two languages, and this might be due to the fact that it is more widespread, especially in the African continent (cf. also Table \ref{tab:breakdown}). 

We bserve a peak of activity across all languages in January, corresponding to the beginning of the vaccination program, and another one in March when alleged links between the AstraZeneca vaccine and blood clots became viral in mainstream media. In summer there is an outstanding increase of French and Italian tweets, probably linked to the introduction of the restrictions for unvaccinated people, whereas towards fall we can see that the topic is trending across all languages (especially German) following a slight increase in the number of vaccinations. In fact, there is a significant Pearson correlation between the daily volumes of tweets collected in different languages (in the range 0.56-0.71, $P\sim0$).

\subsection{Hashtags}


\begin{figure}[!t]
    \centering
    \includegraphics[width=\linewidth]{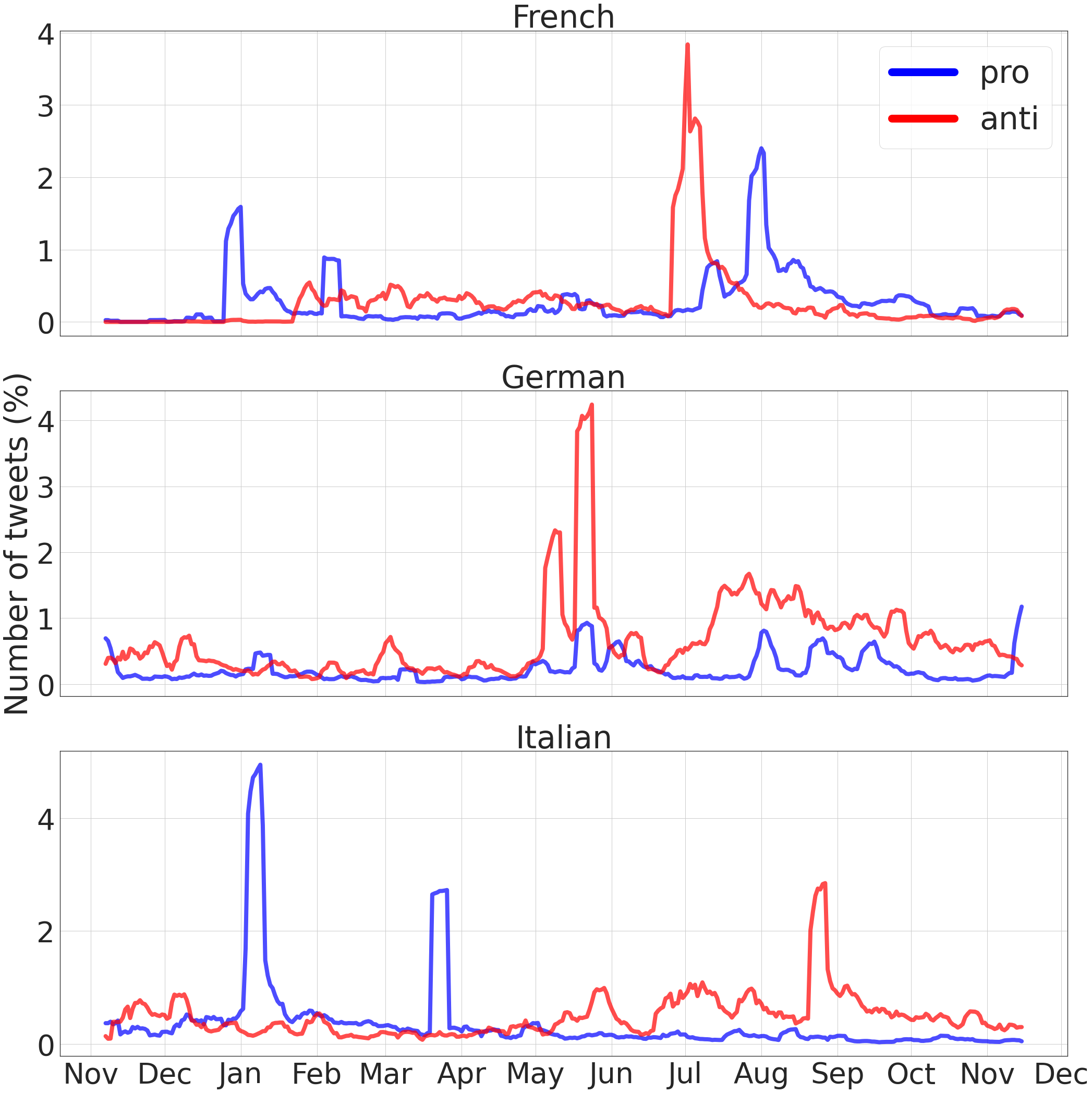}
    \caption{Daily percentage of tweets and retweets sharing pro and anti vaccine hashtags, respectively in blue and red, for each language. We count tweets which contain only hashtags belonging to one of the two classes. Time series are smoothed with a 7-day average.}
    \label{fig:hashtags_ts}
\end{figure}

When we look at the top-10 most shared hashtags in the three languages, we observe that they mostly contain generic references to the pandemic (e.g. "vaccin", "covid19", "corona"), the debate around the introduction of vaccination documents (e.g. "passsanitaire" in French or "greenpass" in Italian) and politicians (e.g. "macron" and "draghi").

In Figure \ref{fig:hashtags_ts} we show instead the daily percentage of tweets sharing Pro and Anti vaccine Gold Hashtags (computed over the total number of tweets shared in that day), using the list of GHs specified in the Data Collection section, for each language. For each day we count tweets and retweets which contain hashtags belonging to only one of the two classes. For what concerns French, we notice a peak of activity for Pro vaccine hashtags at the beginning of the campaign (January 2021) and another in late summer, which follows a strong peak of Anti vaccination hashtags. For what concerns German, we notice little sharing activity for Pro vaccine hashtags, whereas Anti vaccination ones exhibit a peak at the beginning of summer, and then show an increasing trend towards the beginning of fall. Finally, for what concerns Italian, we notice a large number of Italian Pro vaccine hashtags at the beginning of the campaign in January, and likewise in correspondence of the AstraZeneca blood clots 'event'. Towards summer,  similarly to other langauges, we notice an increase in the sharing of Anti vaccination hashtags. Overall, daily volumes stay in a similar range across different languages (0-4\%).

\begin{figure}[!t]
    \centering
    \includegraphics[width=\linewidth]{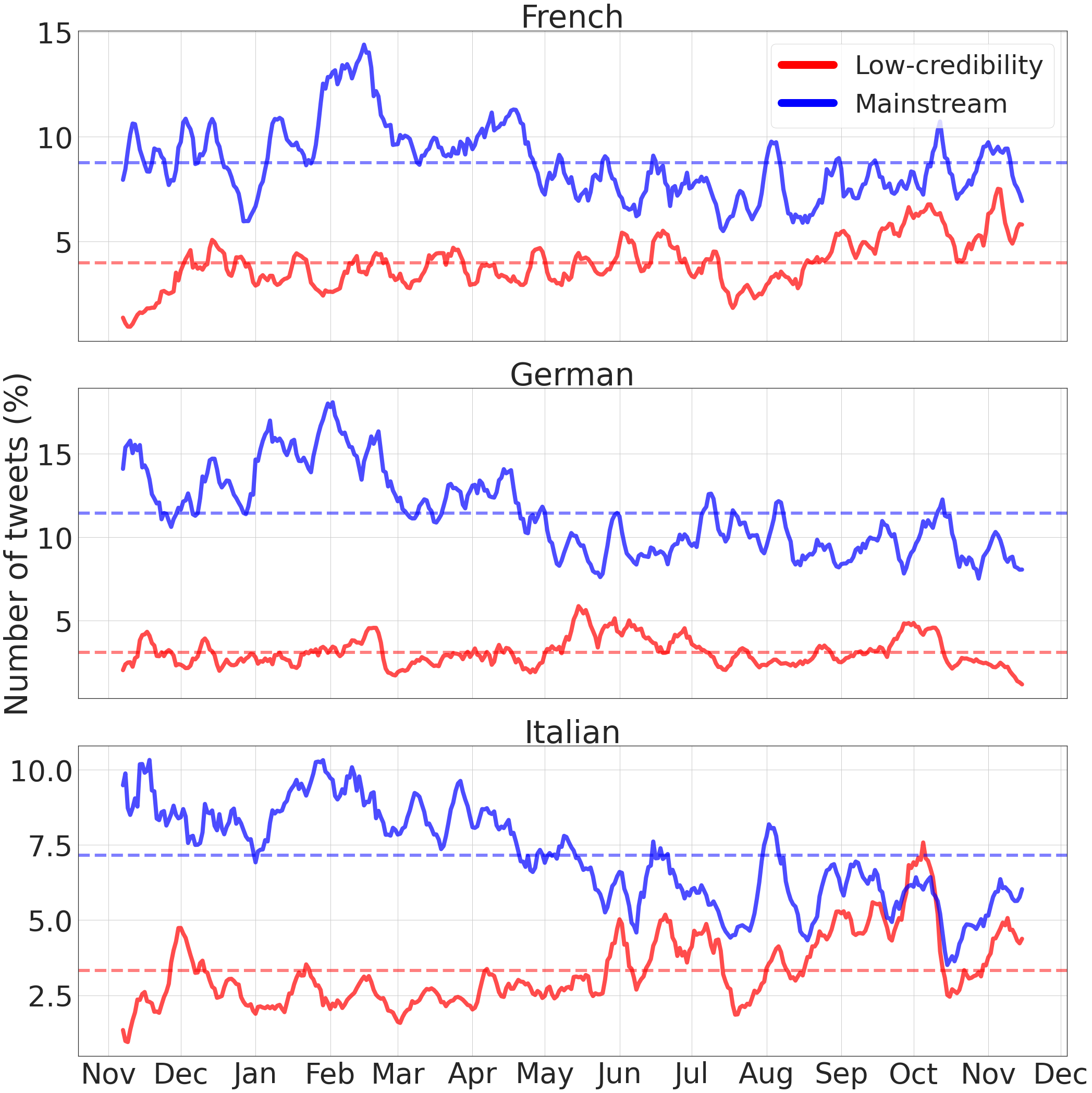}
    \caption{Daily percentage of tweets and retweets sharing links to low-credibility and mainstream news websites, respectively in red and blue, for each language. Time series are smoothed with a 7-day average. Dashed lines represent the mean value over the entire period of observation.}
    \label{fig:low_high_ts}
\end{figure}

\subsection{News sources}
We now investigate the prevalence of low-credibility by using a source-based approach to label news articles, i.e., we label sources based on lists compiled by journalists, researchers and fact-checkers and we propagate the label to all URLs linking to these websites. This approach is limited, since not all stories published on a disinformation website are fake, but it is widely adopted in the literature to study low-credibility content at scale \cite{yang2020covid,Bovet2019,shao2018spread,caldarelli2021flow,brena2019news}. As a reference, we consider publishers of mainstream news as a proxy for reliable information similar to \cite{yang2020covid}.

Specifically, we aggregate three different sources of labels:
\begin{itemize}
    \item a list of 60+ Italian low-credibility websites which were flagged by Italian fact-checkers and journalists for sharing disinformation, misinformation, fake news, etc introduced in \cite{PierriArtoni2020} and employed in \cite{PierriWWW20,Pierri2020epj,guarino2021information, pierri2021vaccinitaly}. It is available in our repository.
    \item a list of over 600 low-credibility domains based on information provided by the Media Bias/Fact Check website (MBFC, \url{mediabiasfactcheck.com}) \cite{yang2020covid}. It is available in our repository.
    \item a list of credibility scores in the range $[0, 100]$ provided by NewsGuard (\url{https://www.newsguardtech.com/it/}), a journalistic organization that rates websites on their tendency to spread true or false information. In particular, we consider publishers with a score less than 60 as low-credibility (as suggested by NewsGuard), and those with a score higher than 60 as mainstream. We cannot disclose this list because the data is proprietary.
\end{itemize}

\begin{figure}[!ht]
    \centering
    \includegraphics[width=0.8\linewidth]{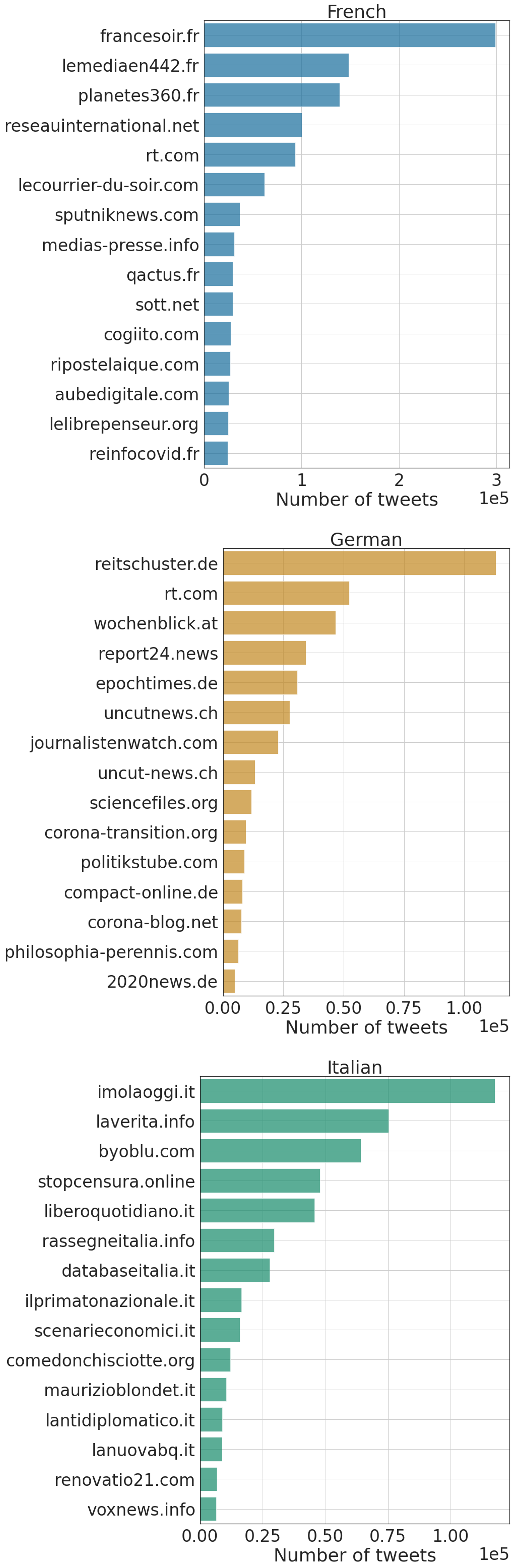}
    \caption{Top-10 most (re)tweeted low-credibility websites in different languages.}
    \label{fig:low_top}
\end{figure}

In Figure \ref{fig:low_high_ts} we show the daily percentage of tweets and retweets containing a link to low-credibility and mainstream news websites. We can see that the amount of low-credibility is smaller yet non negligible compared to mainstream news. It is also stationary around the mean value of the entire period (in the range $[2.5\%, 4.8\%]$) in all languages, whereas mainstream coverage of vaccines exhibits a decreasing trend towards summer for German and Italian. Interestingly, we can notice that around October-November 2021 the amount of Italian misinformation circulating on Twitter was higher than mainstream news. However, we remark that our lists are not exhaustive, and that these estimations should be consider as a lower bound for both low-credibility and mainstream information.

We further investigate which are the most shared low-credibility news websites in different languages. In Figure \ref{fig:low_top} we provide the Top-15 ranking of such websites. We can see a similar prevalence on Twitter of most popular misinformation websites, with the uppermost websites being shared over 100k times. In French: \url{"francesoir.fr"} is a popular tabloid which has been criticised for publishing false claims about the COVID-19 pandemic. In German: \url{"reitschuster.de"} is the blog of a political commentator (Boris Reitschuster) which has a borderline score according to Newsguard (it's rated 59.5 out of 100) and that has been flagged for sharing misinformation about the pandemic. In Italian: \url{"imolaoggi.it"} is a news website which has been repetitiously flagged for sharing hoaxes, misinformation and fake news. We leave further investigation of these websites for future work.

\subsection{Geolocation}
We used the methodology described in \citet{mejova2021} to locate users in our dataset and estimate the geographical composition of the data collected for each language. This employs the GeoNames\footnote{The code for geolocation can be found at \url{https://sites.google.com/site/yelenamejova/resources}} location database to match the user-specified free-text location strings to a location. Not all users can be geolocated in this way, because many do not put a string in the "location" field. We report the following:
\begin{itemize}
    \item French: over 750k users and 17.4 million tweets are geolocated. Around 55\% users are geolocated to France and are responsible for 67\% of the geolocated tweets. Second and third most frequent countries are United States ($\sim7$\% tweets) and Canada ($\sim4$\% tweets).
    \item German: over 270k users and 7.8 million tweets are geolocated. 66\% of the users and tweets are geolocated in Germany. Second and third most frequent countries are Austria ($\sim8$\% tweets) and Switzerland ($\sim7.7$\% tweets).
    \item Italian: over 290k users and 7.5 million tweets are geolocated. Around 52\% of the users are geolocated to Italy, and they shared over 80\% of the geolocated tweets. Second and third most frequent countries are the United States ($\sim 4\%$ tweets) and France ($\sim 3$\% tweets).
\end{itemize}
The approach is not completely accurate, since it is based on a simple string matching, but we can observe that indeed most of the accounts are geolocated in the main countries where each language is spoken, namely France, Germany, and Italy. For what concerns French, we do not get a large number of users geolocated in African countries, but a further investigation is needed to understand whether the geolocation technique is not working properly or Twitter is not very used in those countries.

\subsection{Coordinated activity}
In this section we try to identify coordinated activity on the dataset by applying a coordination detection framework \cite{pacheco2021uncovering}. 
While coordination may occur over many different possible dimensions, here we focus our attention on coordinated sharing of URLs. 
Other dimensions could be explored to identify other coordinated accounts, based for instance on shared hashtag and/or images.

Specifically, for each date in the period under analysis, we built a bipartite network of users and URLs they shared on native tweets (excl. retweets and quote retweets). 
Then, we projected it to users such that two users would be connected if they shared the same URL. 
Edges between users are thus weighted by the number of same URLs that they shared.

To focus on the most suspicious users, we filtered out edges with a weight smaller than 10, and removed singleton nodes resulting from this procedure. 
Finally, we aggregated all daily networks such that edge weights correspond to the number of days in which we found a pair of users sharing the same URLs at least 10 times.

\begin{figure}
    \centering
    \includegraphics[width=0.8\linewidth]{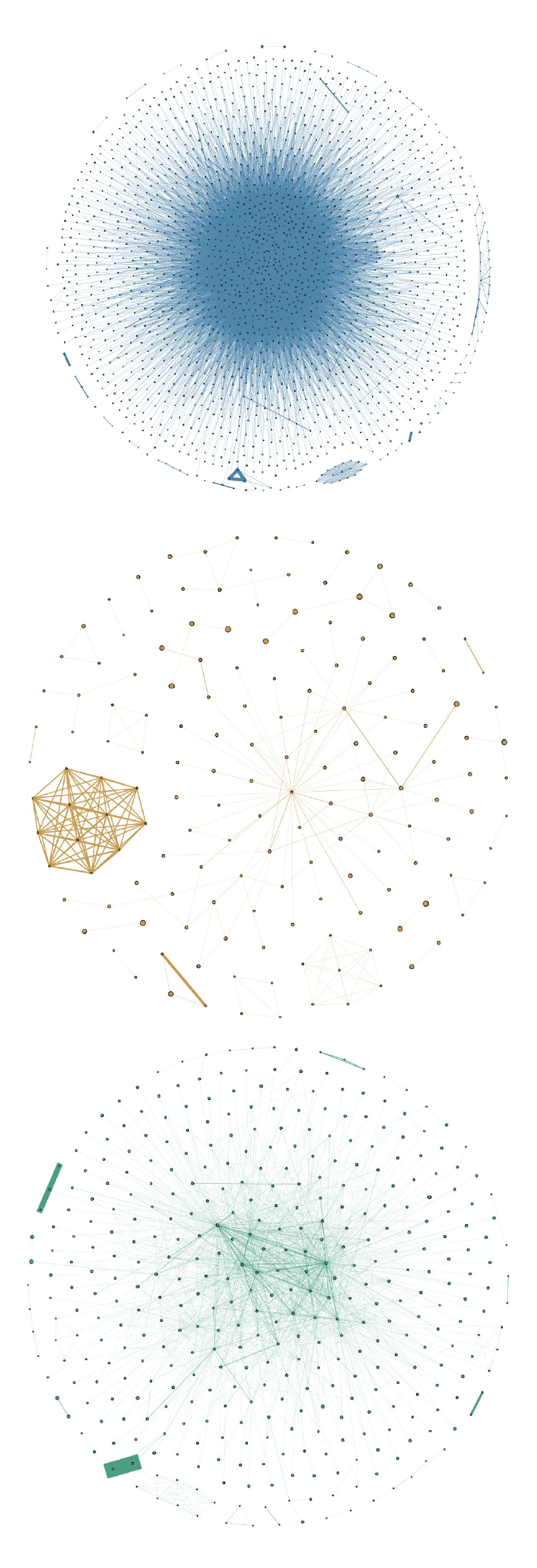}
    \caption{Networks of coordinated accounts that shared the same URLs at least 10 times on a daily basis, based respectively on French (top), German (center) and Italian (bottom) tweets.}
    \label{fig:coordinated}
\end{figure}

The resulting networks, one for each language, can be found in Figure \ref{fig:coordinated}.
The network for French has 1,888 nodes and 28,951 edges, for German it has 157 nodes and 236 edges, and for Italian it has 392 nodes and 1,555 edges. 
The size of nodes corresponds to the percentage of links to low-credibility domains, as defined in the previous section, and edge are ranked by their weight, with thicker edges indicating a higher weight.

The Italian and French networks are dominated by a single large component. 
In contrast, the German one contains two large components. 
Of these two components, the one on the bottom left is densely connected with thicker edges while the one in the middle is sparser with thinner edges. 
This behavior makes the former more suspicious than the latter.
Additionally, all of these components exhibit dissimilarities on uniformity or variety of low credibility sources shared. 
For example, the accounts found in the Italian network shared a lower percentage of these sources compared to those in France network.


\section{Conclusions}
We presented a large-scale dataset of Twitter messages related to vaccines in three different languages (French, German, and Italian), which allows to investigate the impact and the influence of online conversations about COVID-19 vaccines on social media.

We provided a few preliminary analyses of the dataset. We showed that throughout 2021 there were a few peaks of attention around the topic in correspondence of the beginning of vaccination programs, the AstraZeneca blood clots and the introduction of limitations for unvaccinated people. We showed that hashtags expressing positive and negative views about vaccines were highly shared in different periods depending on the language, and that online misinformation accounts for around 5\% of the tweets shared in each language. We also showed that most of the users in our collection reside in three main countries: France, Germany, and Italy. We experimented with a coordinated activity framework highlighting the presence of clusters of users promoting anti-vaccination content in a coordinated fashion.

There a few limitations to our work. First, the procedure used to identify Twitter conversations about COVID-19 vaccines involved a manual evaluation to determine relevant keywords, and thus it might be unable to fully exclude irrelevant data and/or conversations around vaccines which are not COVID-19 specific (e.g. animals, MMR, etc). Still, it allows for further filtering and refinement at a later stage.

Second, Twitter users might not be a representative sample of the population, and their online activity might not reflect the general public opinion \cite{pew2020twitterUsers}. Besides, according to the 2021 Reuters Digital News Report\footnote{\url{https://reutersinstitute.politics.ox.ac.uk/digital-news-report/2021}}, Twitter was used respectively by 17\% of the respondents in France, 6\% in Germany and 8\% in Italy for any purpose. As a matter of fact, Facebook remains the most used social media platform~\cite{boberg2020pandemic} in most countries, but it does not allow to collect relevant data.

Third, users cannot opt-out from our collection, and this might raise important ethical concerns about anonymity. Nevertheless, whenever a user deletes a tweet or account, the related content will be unavailable in the re-hydration process.

There is a number of potential usages for this dataset. We aim to explore the correlation between the prevalence of online misinformation about vaccines \cite{pierri2021impact} and public health outcomes (e.g. COVID-19 vaccine uptake rates, hospitalizations, etc) in different countries. We also plan to further investigate the presence of suspicious accounts, such as bots and trolls, and provide evidence of coordinated campaigns promoting anti-vaccine messages \cite{pacheco2021uncovering}. Finally, we plan to build models to describe how online vaccine misinformation and anti-vaccine sentiment spread in different countries.

\section{Acknowledgments}
This work has been partially supported by the PRIN grant HOPE (FP6, Italian Ministry of Education), and the EU H2020 research and innovation programme, COVID-19 call, under grant agreement No. 101016233 “PERISCOPE” (https://periscopeproject.eu/). We are grateful to Lorenzo Corti, Andrea Tocchetti, Silvio Pavanetto, Pascal Garel, Moritz Laurer, and Anita Gottlob for helping in the selection of relevant keywords, gold hashtags and for manually annotating tweets.

\bibliography{bib}

\begin{thebibliography}{35}
\providecommand{\natexlab}[1]{#1}
\providecommand{\url}[1]{\texttt{#1}}
\providecommand{\urlprefix}{URL }
\expandafter\ifx\csname urlstyle\endcsname\relax
  \providecommand{\doi}[1]{doi:\discretionary{}{}{}#1}\else
  \providecommand{\doi}{doi:\discretionary{}{}{}\begingroup
  \urlstyle{rm}\Url}\fi

\bibitem[{Banda et~al.(2021)Banda, Tekumalla, Wang, Yu, Liu, Ding, Artemova,
  Tutubalina, and Chowell}]{banda2021large}
Banda, J.~M.; Tekumalla, R.; Wang, G.; Yu, J.; Liu, T.; Ding, Y.; Artemova, E.;
  Tutubalina, E.; and Chowell, G. 2021.
\newblock A large-scale COVID-19 Twitter chatter dataset for open scientific
  research—an international collaboration.
\newblock \emph{Epidemiologia} 2(3): 315--324.

\bibitem[{Boberg et~al.(2020)Boberg, Quandt, Schatto-Eckrodt, and
  Frischlich}]{boberg2020pandemic}
Boberg, S.; Quandt, T.; Schatto-Eckrodt, T.; and Frischlich, L. 2020.
\newblock Pandemic populism: Facebook pages of alternative news media and the
  corona crisis--A computational content analysis.
\newblock \emph{arXiv preprint arXiv:2004.02566} .

\bibitem[{Bovet and Makse(2019)}]{Bovet2019}
Bovet, A.; and Makse, H.~A. 2019.
\newblock {Influence of fake news in {Twitter} during the 2016 {US}
  presidential election}.
\newblock \emph{Nature Communications} 10(1): 7.
\newblock ISSN 2041-1723.
\newblock \urlprefix\url{https://doi.org/10.1038/s41467-018-07761-2}.

\bibitem[{Brena et~al.(2019)Brena, Brambilla, Ceri, Di~Giovanni, Pierri, and
  Ramponi}]{brena2019news}
Brena, G.; Brambilla, M.; Ceri, S.; Di~Giovanni, M.; Pierri, F.; and Ramponi,
  G. 2019.
\newblock News sharing user behaviour on twitter: A comprehensive data
  collection of news articles and social interactions.
\newblock In \emph{Proceedings of the International AAAI Conference on Web and
  Social Media}, volume~13, 592--597.

\bibitem[{Broniatowski et~al.(2018)Broniatowski, Jamison, Qi, AlKulaib, Chen,
  Benton, Quinn, and Dredze}]{broniatowski2018weaponized}
Broniatowski, D.~A.; Jamison, A.~M.; Qi, S.; AlKulaib, L.; Chen, T.; Benton,
  A.; Quinn, S.~C.; and Dredze, M. 2018.
\newblock Weaponized health communication: Twitter bots and Russian trolls
  amplify the vaccine debate.
\newblock \emph{American journal of public health} 108(10): 1378--1384.

\bibitem[{Burki(2019)}]{burki2019vaccine}
Burki, T. 2019.
\newblock Vaccine misinformation and social media.
\newblock \emph{The Lancet Digital Health} 1(6): e258--e259.

\bibitem[{Caldarelli et~al.(2021)Caldarelli, De~Nicola, Petrocchi, Pratelli,
  and Saracco}]{caldarelli2021flow}
Caldarelli, G.; De~Nicola, R.; Petrocchi, M.; Pratelli, M.; and Saracco, F.
  2021.
\newblock Flow of online misinformation during the peak of the COVID-19
  pandemic in Italy.
\newblock \emph{EPJ data science} 10(1): 34.

\bibitem[{Chen et~al.(2020)Chen, Lerman, Ferrara et~al.}]{chen2020tracking}
Chen, E.; Lerman, K.; Ferrara, E.; et~al. 2020.
\newblock Tracking social media discourse about the covid-19 pandemic:
  Development of a public coronavirus twitter data set.
\newblock \emph{JMIR Public Health and Surveillance} 6(2): e19273.

\bibitem[{Cinelli et~al.(2021)Cinelli, Morales, Galeazzi, Quattrociocchi, and
  Starnini}]{cinelli2021echo}
Cinelli, M.; Morales, G. D.~F.; Galeazzi, A.; Quattrociocchi, W.; and Starnini,
  M. 2021.
\newblock The echo chamber effect on social media.
\newblock \emph{Proceedings of the National Academy of Sciences} 118(9).

\bibitem[{Cossard et~al.(2020)Cossard, Morales, Kalimeri, Mejova, Paolotti, and
  Starnini}]{cossard_2020}
Cossard, A.; Morales, G. D.~F.; Kalimeri, K.; Mejova, Y.; Paolotti, D.; and
  Starnini, M. 2020.
\newblock Falling into the echo chamber: the Italian vaccination debate on
  Twitter.
\newblock In \emph{Proceedings of the International AAAI Conference on Web and
  Social Media}, volume~14, 130--140.

\bibitem[{DeVerna et~al.(2021)DeVerna, Pierri, Truong, Bollenbacher, Axelrod,
  Loynes, Torres-Lugo, Yang, Menczer, and Bryden}]{deverna2021covaxxy}
DeVerna, M.; Pierri, F.; Truong, B.; Bollenbacher, J.; Axelrod, D.; Loynes, N.;
  Torres-Lugo, C.; Yang, K.-C.; Menczer, F.; and Bryden, J. 2021.
\newblock CoVaxxy: A global collection of English Twitter posts about COVID-19
  vaccines.
\newblock \emph{Proceedings of the International AAAI Conference on Web and
  Social Media} .

\bibitem[{Di~Giovanni et~al.(2021)Di~Giovanni, Corti, Pavanetto, Pierri,
  Tocchetti, and Brambilla}]{digiovanni2021content}
Di~Giovanni, M.; Corti, L.; Pavanetto, S.; Pierri, F.; Tocchetti, A.; and
  Brambilla, M. 2021.
\newblock A Content-based Approach for the Analysis and Classification of
  Vaccine-related Stances on Twitter: the Italian Scenario.
\newblock \emph{Workshop Proceedings of the International AAAI Conference on
  Web and Social Media} .

\bibitem[{Gallotti et~al.(2020)Gallotti, Valle, Castaldo, Sacco, and
  De~Domenico}]{gallotti2020assessing}
Gallotti, R.; Valle, F.; Castaldo, N.; Sacco, P.; and De~Domenico, M. 2020.
\newblock {Assessing the risks of `infodemics' in response to COVID-19
  epidemics}.
\newblock \emph{Nature Human Behaviour} 4: 1285–1293.
\newblock \doi{10.1038/s41562-020-00994-6}.

\bibitem[{Gargiulo et~al.(2020)Gargiulo, Cafiero, Guille-Escuret, Seror, and
  Ward}]{Gargiulo2020Asymmetric}
Gargiulo, F.; Cafiero, F.; Guille-Escuret, P.; Seror, V.; and Ward, J. 2020.
\newblock Asymmetric participation of defenders and critics of vaccines to
  debates on French-speaking Twitter.
\newblock \emph{Scientific Reports} 10.
\newblock \doi{10.1038/s41598-020-62880-5}.

\bibitem[{Guarino et~al.(2021)Guarino, Pierri, Di~Giovanni, and
  Celestini}]{guarino2021information}
Guarino, S.; Pierri, F.; Di~Giovanni, M.; and Celestini, A. 2021.
\newblock Information disorders during the COVID-19 infodemic: The case of
  Italian Facebook.
\newblock \emph{Online Social Networks and Media} 22: 100124.

\bibitem[{Hayawi et~al.(2021)Hayawi, Shahriar, Serhani, Taleb, and
  Mathew}]{HAYAWI2021}
Hayawi, K.; Shahriar, S.; Serhani, M.~A.; Taleb, I.; and Mathew, S.~S. 2021.
\newblock ANTi-Vax: A Novel Twitter Dataset for COVID-19 Vaccine Misinformation
  Detection.
\newblock \emph{Public Health} ISSN 0033-3506.
\newblock \doi{https://doi.org/10.1016/j.puhe.2021.11.022}.
\newblock
  \urlprefix\url{https://www.sciencedirect.com/science/article/pii/S0033350621004534}.

\bibitem[{Imran, Qazi, and Ofli(2021)}]{imran2021tbcov}
Imran, M.; Qazi, U.; and Ofli, F. 2021.
\newblock TBCOV: Two Billion Multilingual COVID-19 Tweets with Sentiment,
  Entity, Geo, and Gender Labels.
\newblock \emph{arXiv preprint arXiv:2110.03664 - Forthcoming in Data} .

\bibitem[{Johnson et~al.(2020)Johnson, Vel{\'a}squez, Restrepo, Leahy, Gabriel,
  El~Oud, Zheng, Manrique, Wuchty, and Lupu}]{johnson2020online}
Johnson, N.~F.; Vel{\'a}squez, N.; Restrepo, N.~J.; Leahy, R.; Gabriel, N.;
  El~Oud, S.; Zheng, M.; Manrique, P.; Wuchty, S.; and Lupu, Y. 2020.
\newblock The online competition between pro-and anti-vaccination views.
\newblock \emph{Nature} 1--4.

\bibitem[{Loomba et~al.(2021)Loomba, de~Figueiredo, Piatek, de~Graaf, and
  Larson}]{loomba2021measuring}
Loomba, S.; de~Figueiredo, A.; Piatek, S.~J.; de~Graaf, K.; and Larson, H.~J.
  2021.
\newblock Measuring the impact of COVID-19 vaccine misinformation on
  vaccination intent in the UK and USA.
\newblock \emph{Nature human behaviour} 5(3): 337--348.

\bibitem[{Lopez and Gallemore(2021)}]{lopez2021augmented}
Lopez, C.~E.; and Gallemore, C. 2021.
\newblock An augmented multilingual Twitter dataset for studying the COVID-19
  infodemic.
\newblock \emph{Social Network Analysis and Mining} 11(1): 1--14.

\bibitem[{Mathieu et~al.(2021)Mathieu, Ritchie, Ortiz-Ospina, Roser, Hasell,
  Appel, Giattino, and Rod{\'e}s-Guirao}]{owid}
Mathieu, E.; Ritchie, H.; Ortiz-Ospina, E.; Roser, M.; Hasell, J.; Appel, C.;
  Giattino, C.; and Rod{\'e}s-Guirao, L. 2021.
\newblock A global database of COVID-19 vaccinations.
\newblock \emph{Nature human behaviour} 1--7.

\bibitem[{Mejova and Kourtellis(2021)}]{mejova2021}
Mejova, Y.; and Kourtellis, N. 2021.
\newblock YouTubing at Home: Media Sharing Behavior Change as Proxy for
  Mobility Around COVID-19 Lockdowns.
\newblock In \emph{13th ACM Web Science Conference 2021}, WebSci '21,
  272–281. New York, NY, USA: Association for Computing Machinery.
\newblock ISBN 9781450383301.
\newblock \doi{10.1145/3447535.3462494}.
\newblock \urlprefix\url{https://doi.org/10.1145/3447535.3462494}.

\bibitem[{Muric, Wu, and Ferrara(2021)}]{muric2021}
Muric, G.; Wu, Y.; and Ferrara, E. 2021.
\newblock COVID-19 Vaccine Hesitancy on Social Media: Building a Public Twitter
  Data Set of Antivaccine Content, Vaccine Misinformation, and Conspiracies.
\newblock \emph{JMIR Public Health Surveill} 7(11): e30642.
\newblock ISSN 2369-2960.
\newblock \doi{10.2196/30642}.
\newblock \urlprefix\url{https://publichealth.jmir.org/2021/11/e30642}.

\bibitem[{Orenstein and Ahmed(2017)}]{orenstein2017simply}
Orenstein, W.~A.; and Ahmed, R. 2017.
\newblock Simply put: Vaccination saves lives.
\newblock \emph{Proceedings of the National Academy of Sciences} 114(16):
  4031--4033.

\bibitem[{Pacheco et~al.(2021)Pacheco, Hui, Torres-Lugo, Truong, Flammini, and
  Menczer}]{pacheco2021uncovering}
Pacheco, D.; Hui, P.-M.; Torres-Lugo, C.; Truong, B.~T.; Flammini, A.; and
  Menczer, F. 2021.
\newblock Uncovering Coordinated Networks on Social Media: Methods and Case
  Studies.
\newblock In \emph{Proceedings of the International AAAI Conference on Web and
  Social Media}, volume~15, 455--466.

\bibitem[{Pierri(2020)}]{PierriWWW20}
Pierri, F. 2020.
\newblock The diffusion of mainstream and disinformation news on Twitter: the
  case of Italy and France.
\newblock \emph{Companion Proceedings of the Web Conference 2020 (WWW ’20
  Companion)} .

\bibitem[{Pierri, Artoni, and Ceri(2020)}]{PierriArtoni2020}
Pierri, F.; Artoni, A.; and Ceri, S. 2020.
\newblock Investigating Italian disinformation spreading on Twitter in the
  context of 2019 European elections.
\newblock \emph{PloS one} 15(1): e0227821.

\bibitem[{Pierri et~al.(2021{\natexlab{a}})Pierri, Perry, DeVerna, Yang,
  Flammini, Menczer, and Bryden}]{pierri2021impact}
Pierri, F.; Perry, B.; DeVerna, M.~R.; Yang, K.-C.; Flammini, A.; Menczer, F.;
  and Bryden, J. 2021{\natexlab{a}}.
\newblock The impact of online misinformation on US COVID-19 vaccinations.
\newblock \emph{arXiv preprint arXiv:2104.10635} .

\bibitem[{Pierri, Piccardi, and Ceri(2020)}]{Pierri2020epj}
Pierri, F.; Piccardi, C.; and Ceri, S. 2020.
\newblock A multi-layer approach to disinformation detection in US and Italian
  news spreading on Twitter.
\newblock \emph{EPJ Data Science} 9(35).

\bibitem[{Pierri et~al.(2021{\natexlab{b}})Pierri, Tocchetti, Corti, Giovanni,
  Pavanetto, Brambilla, and Ceri}]{pierri2021vaccinitaly}
Pierri, F.; Tocchetti, A.; Corti, L.; Giovanni, M.; Pavanetto, S.; Brambilla,
  M.; and Ceri, S. 2021{\natexlab{b}}.
\newblock VaccinItaly: monitoring Italian conversations around vaccines on
  Twitter and Facebook.
\newblock \emph{Workshop Proceedings of the International AAAI Conference on
  Web and Social Media} .

\bibitem[{Righetti(2020)}]{righetti_2020}
Righetti, N. 2020.
\newblock Health Politicization and Misinformation on Twitter. A Study of the
  Italian Twittersphere from Before, During and After the Law on Mandatory
  Vaccinations.
\newblock \doi{10.31219/osf.io/6r95n}.
\newblock \urlprefix\url{osf.io/6r95n}.

\bibitem[{Shao et~al.(2018)Shao, Ciampaglia, Varol, Yang, Flammini, and
  Menczer}]{shao2018spread}
Shao, C.; Ciampaglia, G.~L.; Varol, O.; Yang, K.-C.; Flammini, A.; and Menczer,
  F. 2018.
\newblock The spread of low-credibility content by social bots.
\newblock \emph{Nature Communications} 9: 4787.
\newblock \doi{10.1038/s41467-018-06930-7}.

\bibitem[{Wojick and Hughes(2020)}]{pew2020twitterUsers}
Wojick, S.; and Hughes, A. 2020.
\newblock Sizing Up Twitter Users.
\newblock Pew Research Center,
  \url{https://www.pewresearch.org/internet/2019/04/24/sizing-up-twitter-users/}
  (accessed January, 2021).

\bibitem[{Yang et~al.(2021)Yang, Pierri, Hui, Axelrod, Torres-Lugo, Bryden, and
  Menczer}]{yang2020covid}
Yang, K.-C.; Pierri, F.; Hui, P.-M.; Axelrod, D.; Torres-Lugo, C.; Bryden, J.;
  and Menczer, F. 2021.
\newblock The COVID-19 Infodemic: Twitter versus Facebook.
\newblock \emph{{Big Data \& Society}} Special issue "Studying the COVID-19
  Infodemic at Scale".

\bibitem[{Zarocostas(2020)}]{zarocostas2020fight}
Zarocostas, J. 2020.
\newblock How to fight an infodemic.
\newblock \emph{The Lancet} 395(10225): 676.

\end{thebibliography}

\end{document}